\documentclass[amssymb,amsmath,aps,showpacs]{revtex4}
\usepackage{amssymb}
\usepackage{graphicx}
\usepackage{color}
\usepackage{soul}
\usepackage{latexsym}

\newcommand{\fslash}[1]{{#1 \kern -0.7em/ \kern 0.1em}}

\begin{document}

\voffset 1.25cm

\title{Dark Matter as a Possible New Energy Source for Future Rocket Technology}
\author{Jia Liu }
\email{jl3473@nyu.edu} \affiliation{ Center for Cosmology and
Particle Physics, Department
of Physics, New York University, New York, NY 10003, USA \\
 Institute of Theoretical Physics, School of Physics, Peking University, Beijing 100871,
P.R. China}

\date{\today}

\begin{abstract}
Current rocket technology can not send the spaceship very far,
because the amount of the chemical fuel it can take is limited. We
try to use dark matter (DM) as fuel to solve this problem. In this
work, we give an example of DM engine using dark matter annihilation
products as propulsion. The acceleration is proportional to the
velocity, which makes the velocity increase exponentially with time
in non-relativistic region. The important points for the
acceleration are how dense is the DM density and how large is the
saturation region. The parameters of the spaceship may also have
great influence on the results. We show that the (sub)halos can
accelerate the spaceship to velocity $ 10^{ - 5} c \sim 10^{ - 3}
c$. Moreover, in case there is a central black hole in the halo,
like the galactic center, the radius of the dense spike can be large
enough to accelerate the spaceship close to the speed of light.
\end{abstract}



\maketitle

\section{Introduction}
It is difficult for human to reach the stars using current rocket
technology. The energy sources range from chemical fuel, nuclear
power and even anti-matter conceptually. The major problem in these
systems is that propulsion required large amount of time and
fuel\cite{Lemos}. We try to solve this problem by starting at the
requirement of large amount of fuel. We all know that current
rockets in function are chemical rockets which take oxidant and fuel
at the same time. Interestingly, the airplanes with similar
propulsion system only take fuel, without oxidant, because in the
atmosphere there are enough oxygen which are absorbed by airplane
engines during the flight. Similarly, if there are enough fuel in
the universe, the spaceship may absorb them during its flight like
airplanes absorb the oxygen. Fortunately, dark matter is widely
spread in the universe and the mass density is about five times of
the baryonic matter density\cite{Komatsu:2008hk}, which make it a
possible new energy source for interstellar flight. Thus the
requirement of fuel may be solved in the self-help way with dark
matter as the energy source.

\section{DM engine and acceleration in the saturation density}
\begin{figure}[h]
\vspace*{-.03in} \centering
\includegraphics[width=3.0in,angle=0]{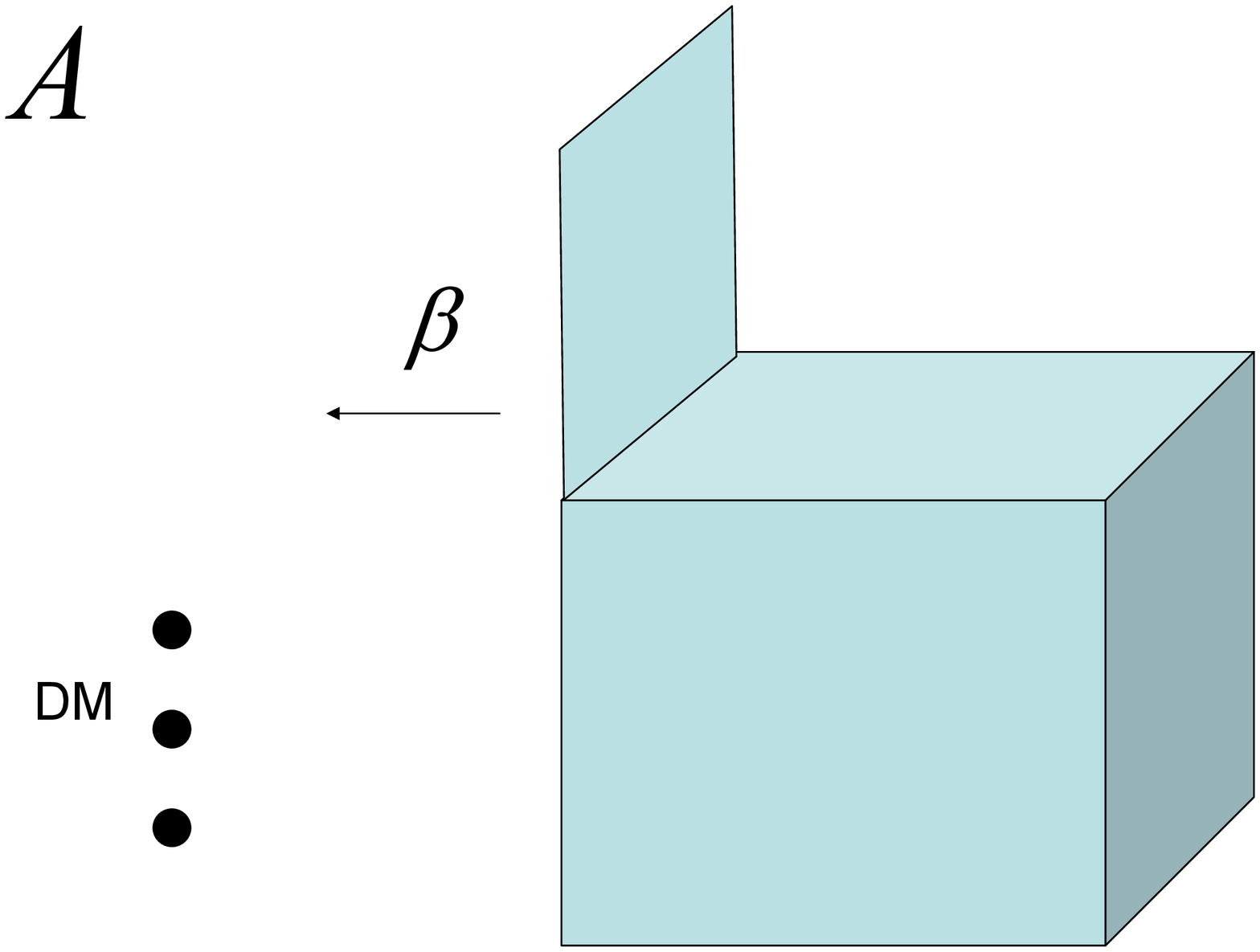}%
\includegraphics[width=3.0in,angle=0]{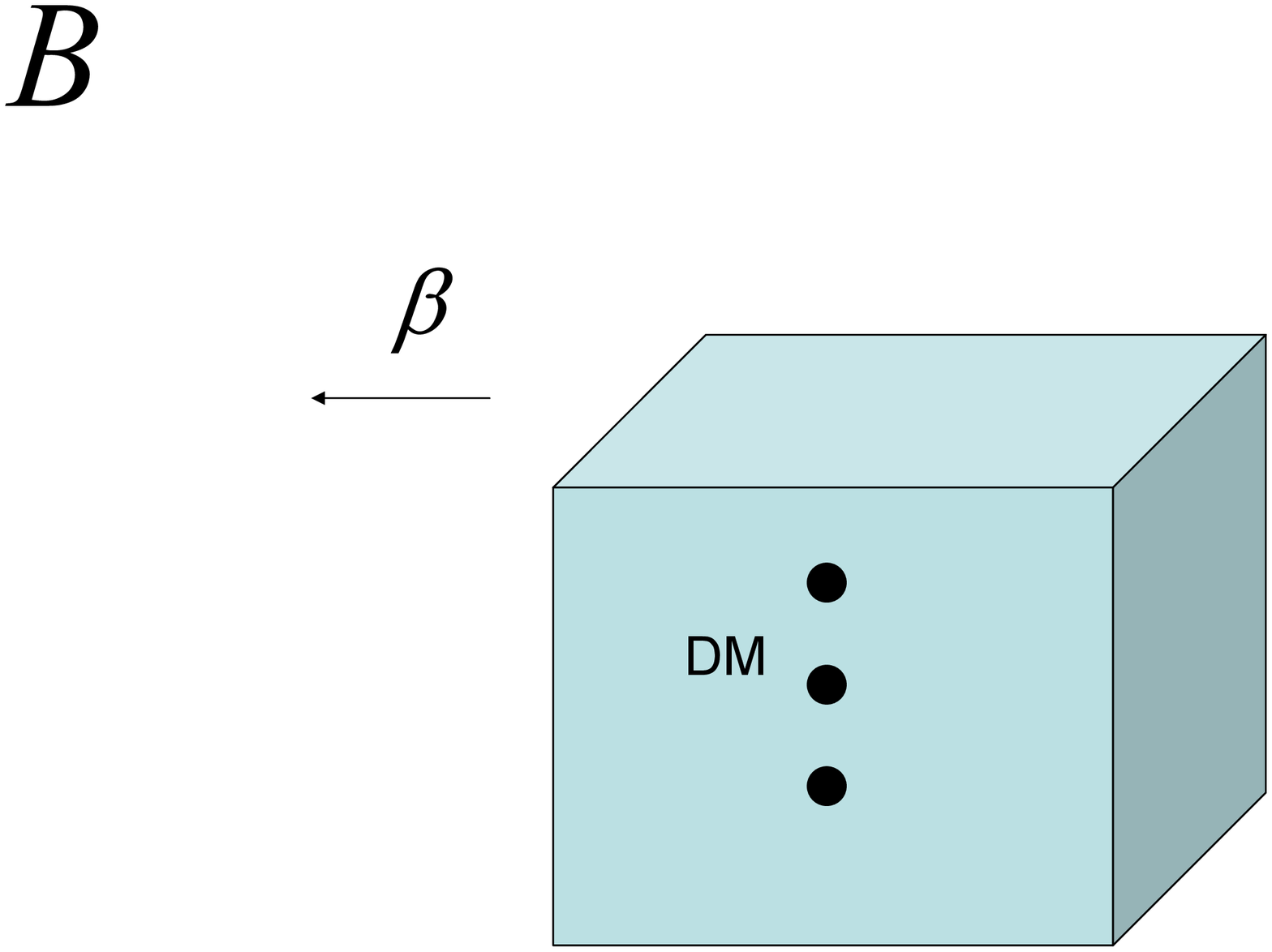}
\\
\includegraphics[width=3.0in,angle=0]{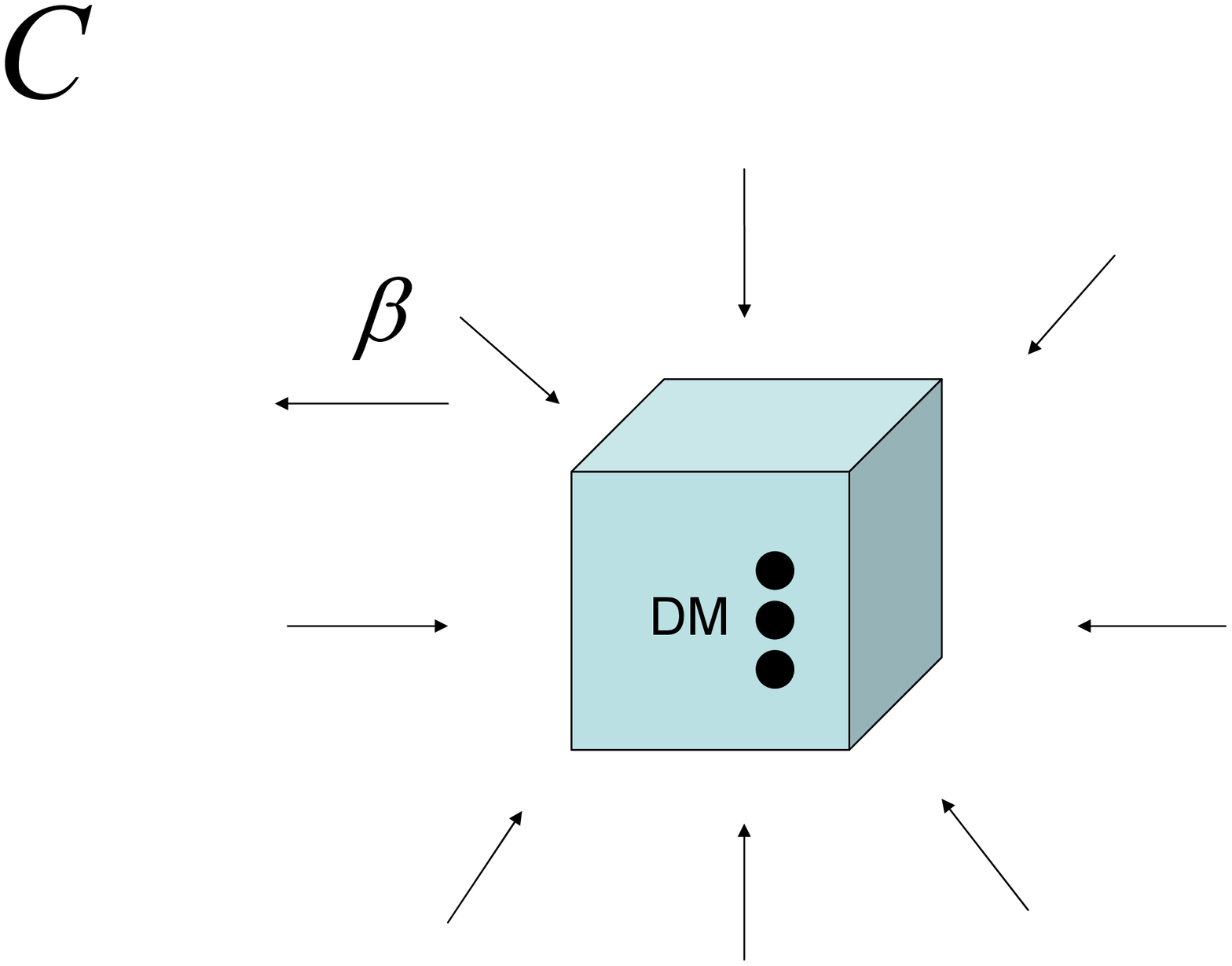}%
\includegraphics[width=3.0in,angle=0]{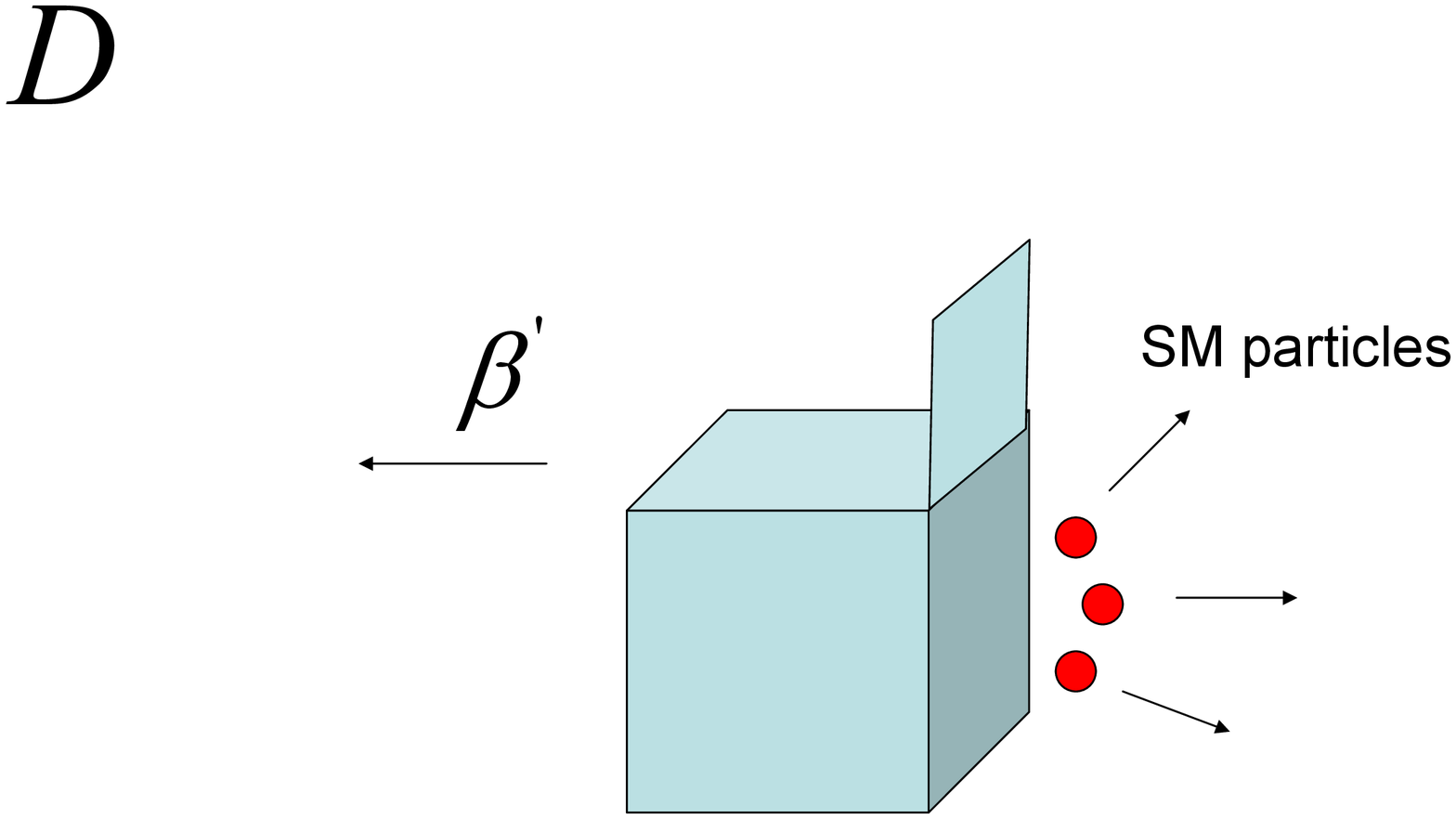}
\vspace*{-.03in} \caption{The illustration of work cycle for the DM
engine.
 \vspace*{-.1in}}
\label{cartoon}
\end{figure}

We give an example of DM engine which uses DM annihilation remnants
as propulsion. Fig.\ref{cartoon} is a sketch of the DM engine for
this kind of new spaceship. The DM engine is the box in the picture.
Here we assume the DM particle and the annihilation products can not
pass through the wall of the box. In picture A, the space ship moves
very fast from right to left. The DM particles, which are assumed to
be static, go into the box and are absorbed in the picture B. In the
picture C, we compress the box and raise the number density of the
DM for annihilation, where we assume the annihilation process
happens immediately. In the picture D, only the wall on the right
side is open. The annihilation products, for example Standard Model
(SM) particles, are all going to the right direction. The processes
from A to D are the working cycle for the engine. Thus, the
spaceship is boosted by the recoil of these SM particles. Note the
spaceship can decelerate by the same system when it reaches the
destination, by opening the left wall in the picture D.

This kind of new spaceship has a very interesting character that the
faster it is, the easier it accelerates. In the picture A, we assume
the rest mass of the spaceship is $ M$ and its velocity is $\beta$
in the unit of speed of light. The time for one cycle of the engine
is $ dt $ and the area of the engine is $ S$. During one work cycle,
the number of DM particles collected by the engine is $ N = \beta dt
\cdot S \cdot \frac{{\rho _D }}{{m_D }}$ , where $ {\rho _D }$ and $
{m_D }$ are the density of DM and mass of DM, respectively. In
picture D, we assume there are only one kind of particles X as the
annihilation products for simplicity. The annihilation process is $
DD \to X\bar X$, with the mass $ m_X < m_D$. For the DM particles,
we assume DM mass $ m_D \sim O(100GeV)$ and the annihilation
products to be SM fermions mainly, which are quite natural in
SuperSymmetry and Extra Dimension models. Thus, the mass of the
annihilation products $m_X$ are quite small comparing with the mass
of dark matter $m_D$. So it is reasonable to use the approximation $
m_X = 0$, where products are treated as massless photon in the
following calculation. Using the conservation of energy and
momentum, we can get

\begin{equation}
\left\{ {\begin{array}{*{20}c}
   {\beta dt \cdot S \cdot \rho _D  + p^0  = p^0  + dp^0  + \varepsilon }  \\
   {0 + p^1  = p^1  + dp^1  - \varepsilon  \cdot \theta }  \\
\end{array}} \right.  \label{conservation}
\end{equation}

where $p^0  = M\gamma $ and $ p^1  = M\gamma \beta $ are the energy
and the momentum of the spaceship, and $ \gamma  \equiv (1 - \beta
^2 )^{ - 1/2}$ , $ \varepsilon$ is the energy of the massless
photons. $ \theta$ is defined as the propulsion efficiency, which
means $ \theta  \in [0,1]$. For example, if the annihilation
particles all go to the right direction, then $ \theta = 1$.
However, if the annihilation particles have equal possibility go
into any direction in the right hemisphere, then $ \theta = 1/2$.
Moreover, it can also be used to count in the other inefficiency of
the engine. By the Eqn.\ref{conservation}, one can get the
differential equation for the velocity,

\begin{equation}
\frac{{k\beta }}{{\theta ^{ - 1}  + \beta }} = \gamma ^3
\frac{{d\beta }}{{dt}}  \label{differential}
\end{equation}

where $ k \equiv S\rho _D /M$. In the non-relativistic region, the
above equation has the simple form,

\begin{equation}
\theta k \cdot \beta  = \frac{{d\beta }}{{dt}}.
\end{equation}

To carry out the numerical calculations, we give some reasonable
parameters first. We assume the weight of the spaceship is $M =
100ton$ and the area is $S = 100m^2$, according to the current
rockets and space shuttles. For the DM density, we use the
saturation density in the center of cusped halos $ \rho _{sat}$. The
saturation density in the halo is due to the balance between the
annihilation rate of the DM $ [\left\langle {\sigma v} \right\rangle
\rho _{sat} /m_D ]^{ - 1}$ and the gravitational in falling rate of
the DM $ (G\bar \rho )^{ - 1/2}$, where the $ {\bar \rho }$ is taken
to be $200$ times of the critical density. Thus the saturation
density is $ \rho _{sat} \sim 10^{19} M_ \odot   \cdot kpc^{ - 3}
$\cite{Berezinsky:1992mx, Lavalle:1900wn}. The propulsion efficiency
is taken to be $ \theta  = 0.5$, since in the picture D we assume
the annihilation particles have equal possibility go into any
direction in the right hemisphere. We can also rewrite the parameter
$k$ as following,

\begin{equation}
k = 2 \times 10^{ - 4} s^{ - 1}  \cdot \left( {\frac{\rho }{{10^{19}
M_ \odot   \cdot kpc^{ - 3} }}\frac{S}{{100m^2 }}\frac{{100ton}}{M}}
\right)
\end{equation}

\begin{figure}[h]
\vspace*{-.03in} \centering
\includegraphics[width=3.4in,angle=0]{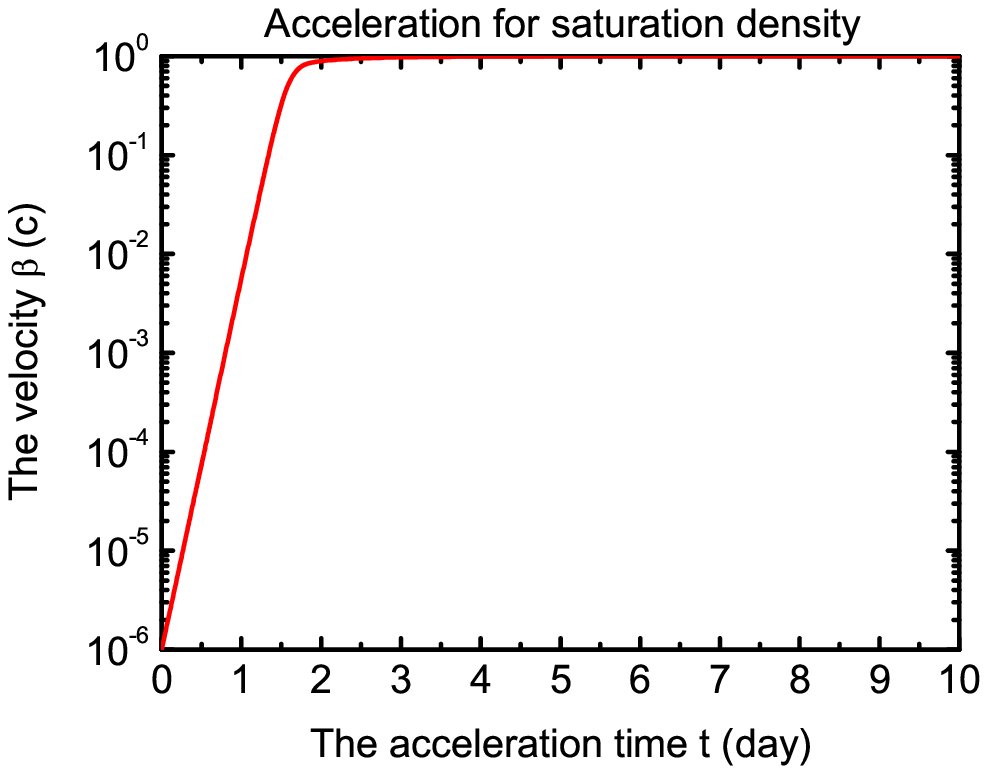}%
\includegraphics[width=3.4in,angle=0]{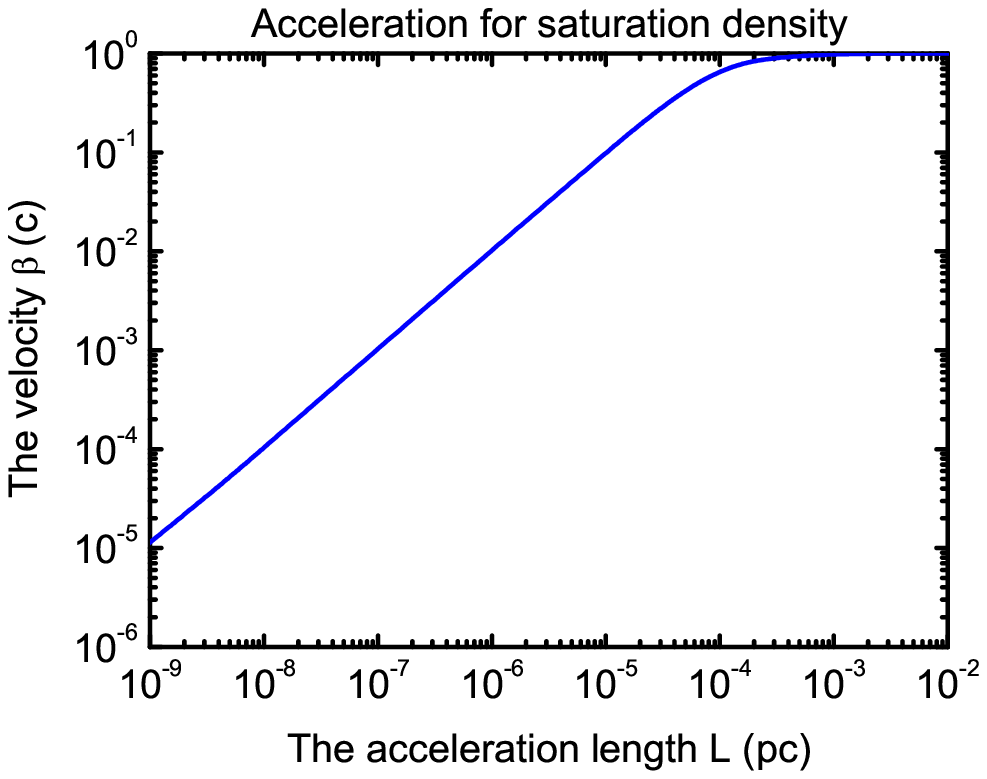}
\vspace*{-.03in} \caption{The velocity as a function of the time and
length needed for the acceleration in the saturation density.
 \vspace*{-.1in}}
\label{constdensity}
\end{figure}

One can solve the Eqn.\ref{differential} and get the time and length
needed for acceleration as the function of velocity,

\begin{equation}
t = (\theta k)^{ - 1}  \cdot \left. {[\frac{{1 + \theta \beta
}}{{\sqrt {1 - \beta ^2 } }} + In(\frac{\beta }{{1 + \sqrt {1 -
\beta ^2 } }})]} \right|_{\beta _0 }^\beta
 \label{contime},
\end{equation}

\begin{equation}
L = (\theta k)^{ - 1}  \cdot \left. {\frac{{\beta  + \theta
}}{{\sqrt {1 - \beta ^2 } }}} \right|_{\beta _0 }^\beta.
 \label{conlength}
\end{equation}

where $ {\beta _0 }$ is the initial velocity at the $ t = 0$. We
give the plots of the above equations in Fig.\ref{constdensity}. We
can see the velocity increases exponentially with time, since the
acceleration is proportional to the velocity. In the
non-relativistic region, where $ \beta  \ll 1$, the
Eqn.\ref{contime} and the Eqn.\ref{conlength} can be simplified as

\begin{equation}
\beta  = \beta _0 e^{\theta kt} ,
\end{equation}

\begin{equation}
L = (\theta k)^{ - 1} (\beta  - \beta _0 ) . \label{conlength1}
\end{equation}

The initial velocity $ {\beta _0 }$ is taken to be $ 10^{ - 6} c$
which is much smaller than the first cosmic velocity. However, the
result is not sensitive to the initial velocity, because of the
exponential increase. In Fig.\ref{constdensity}, we see the
spaceship can reach the relativistic speed in about $2$ $days$ and
the length needed for acceleration is about $ 10^{ - 4} pc$. From
the above equations, if the DM density $ \rho$
 and the area of the spaceship $S$ are
larger, the time $t$ and length $L$ needed for acceleration will go
down. If the mass of spaceship $M$ is larger, the time $t$ and
length $L$ needed for acceleration will increase. However, the mass
of DM particle does not have great influence on results, but the DM
density does.

\section{Acceleration in the halo or subhalo}

Before celebration for the reach of relativistic speed, we should
check whether the saturation region in the halo or subhalo is large
enough for the above calculation. The DM profile can be
parameterized as
$\rho(r)=\frac{\rho_s}{(r/r_s)^{\gamma}[1+(r/r_s)^{\alpha}]^{(\beta-\gamma)/\alpha}}$,
where $\rho_s$ and $r_s$ are the scale density and scale radius
parameters respectively. The parameters $(\alpha,\beta,\gamma)$ are
$(1,3,1)$ for NFW profile\cite{Navarro:1996gj}. Since we are
interested in the central region of halo, where $ r \ll  r_s$, the
profile can be simplified as,

\begin{equation}
\rho  = \frac{{\rho _s r_s }}{r}. \label{NFW}
\end{equation}

\begin{figure}[h]
\vspace*{-.03in} \centering
\includegraphics[width=3.4in,angle=0]{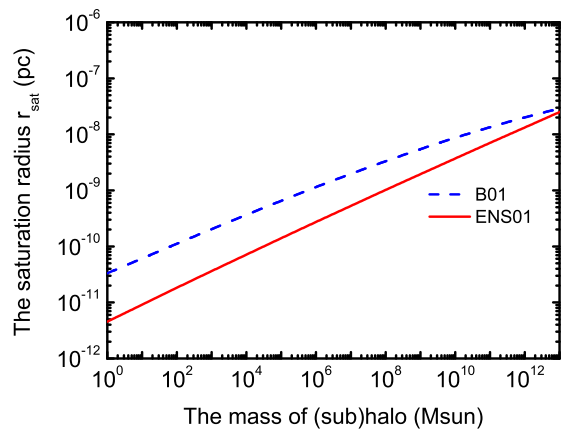}
\vspace*{-.03in} \caption{The saturation radius $ r_{sat}$ for
different (sub)halo mass and concentration models. The B01 and ENS01
stand for different concentration models which are described in the
appendix.
 \vspace*{-.1in}}
\label{rsat}
\end{figure}

This profile is singular at the center of the halo. It is natural to
have cut-off for this singularity due to the balance between the
annihilation rate of the DM and the gravitational in falling rate of
the DM. The saturation DM density is taken to be $ \rho _{sat}  \sim
10^{19} M_ \odot   \cdot kpc^{ - 3} $, thus the radius of saturation
is $ r_{sat}  = {{\rho _s r_s } \mathord{\left/
 {\vphantom {{\rho _s r_s } {\rho _{sat} }}} \right.
 \kern-\nulldelimiterspace} {\rho _{sat} }}$.
 Once we know the scale density $\rho_s$ and scale radius $r_s$,
we can calculate the saturation radius $ r_{sat}$. The $\rho_s$ and
$r_s$ can be fully determined by the concentration model and DM halo
mass, which will be calculated in the appendix. Here we show the
saturation radius $ r_{sat}$ in Fig.\ref{rsat}. We can see the
saturation radius of halo or subhalo is much smaller than the
required length for acceleration to the relativistic speed.

\begin{figure}[h]
\vspace*{-.03in} \centering
\includegraphics[width=3.4in,angle=0]{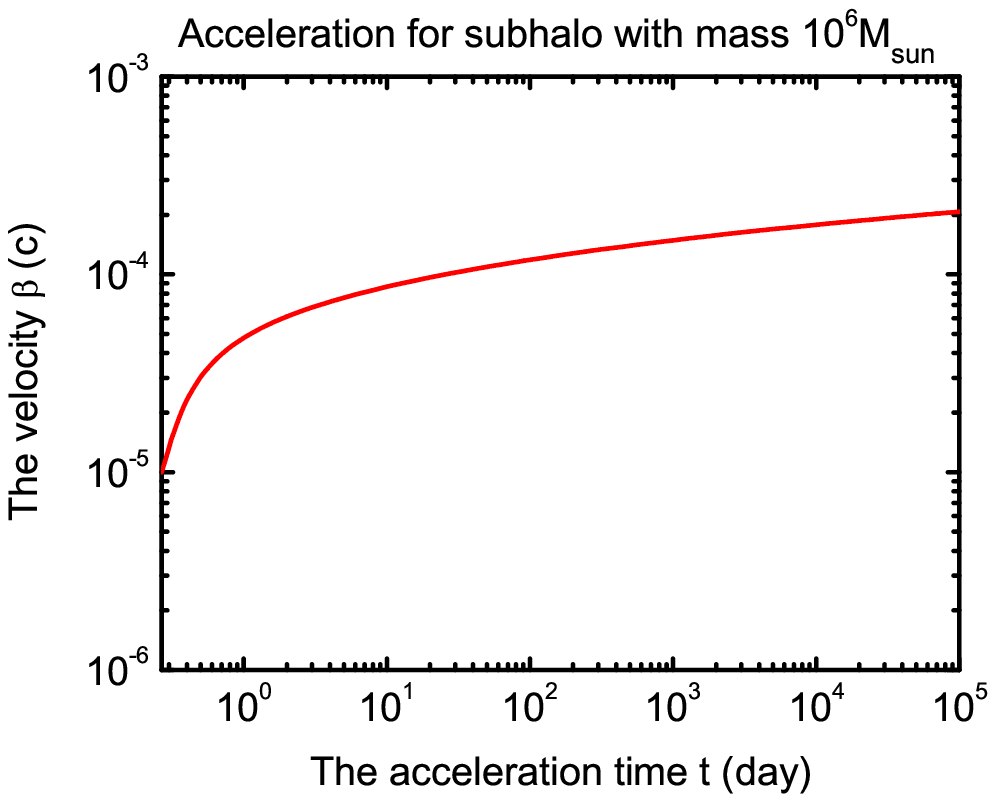}%
\includegraphics[width=3.4in,angle=0]{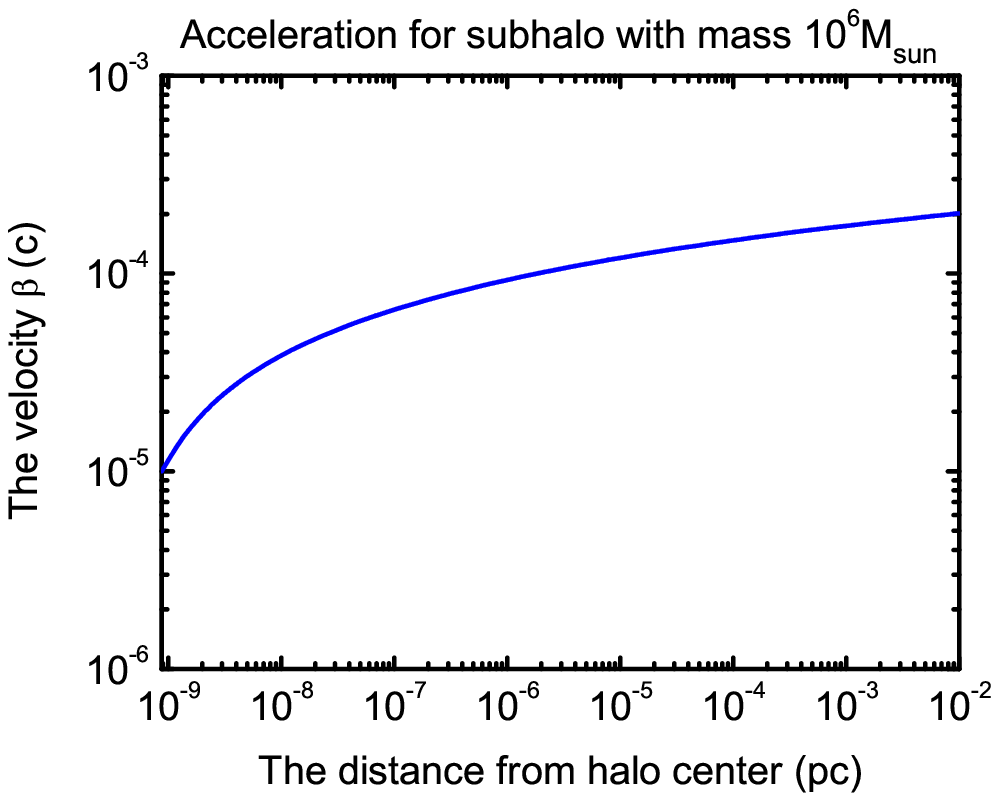}
\vspace*{-.03in} \caption{The velocity as a function of the time and
length needed for the acceleration in the DM (sub)halo. The
spaceship starts from the (sub)halo center with initial velocity $
\beta _0  = 10^{ - 6}c$.
 \vspace*{-.1in}}
\label{real}
\end{figure}

In Fig.\ref{real}, we show the details of acceleration in the
subhalo. The subhalo with mass $ 10^6 M_ \odot$ in B01 model is
taken as an example, which has saturation radius of about $ 10^{ -
9} pc$. Starting from the center of subhalo with initial velocity $
\beta _0 = 10^{ - 6} c$, it reaches the velocity of about $ 10^{ -
5} c$ when it leaves the saturation region, which can be read out
from the Fig.\ref{constdensity}. However, the rest of the subhalo is
not sufficient to accelerate the spaceship to the relativistic
speed, since the density begins to decrease by $ r^{ - 1}$. By
solving the differential equation numerically, we can get the
relations among velocity, time and distance in Fig.\ref{real}. We
can see the spaceship reaches the velocity $ 10^{ - 4} c$ in about
ten days. However, its velocity hardly increases after that, since
the DM density goes down quickly. We can see that the acceleration
is fastest in the saturation area of the halo. But the rest region
of subhalo can still accelerate the spaceship from the velocity $
10^{ - 5} c$ to the velocity $ 10^{ - 4} c$.

\begin{figure}[h]
\vspace*{-.03in} \centering
\includegraphics[width=3.4in,angle=0]{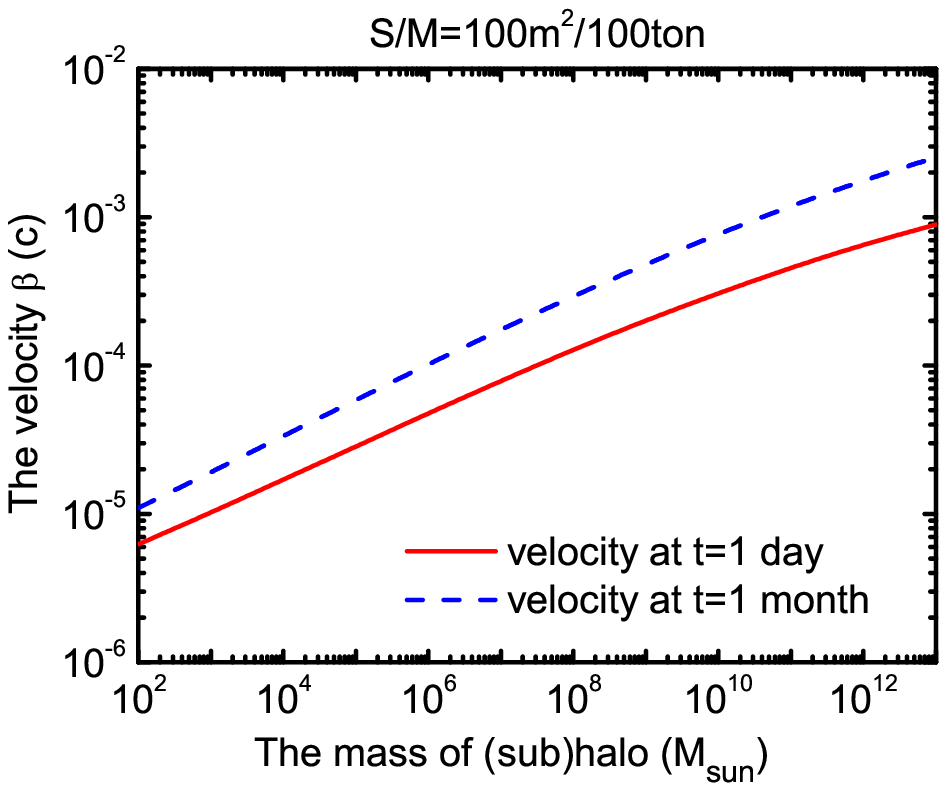}%
\includegraphics[width=3.4in,angle=0]{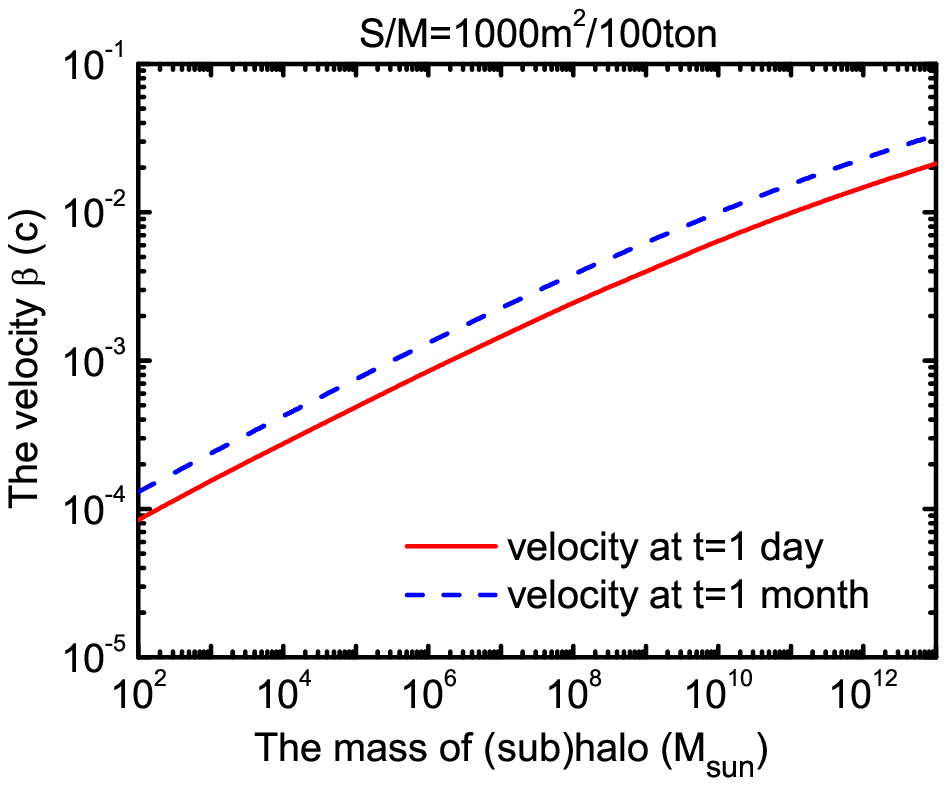}
\vspace*{-.03in} \caption{The velocity at time $t=1 day$ and $t=1
month$ for different (sub)halos mass and spaceship parameters. The
concentration model is taken to be B01. The spaceship is still
assumed to be started at the center of (sub)halo with initial
velocity $ \beta _0 = 10^{ - 6} c$.
 \vspace*{-.1in}}
\label{subhalov}
\end{figure}

To better understand the acceleration power of the (sub)halos, we
give the velocity at different times for different (sub)halos mass
and spaceship parameters in Fig.\ref{subhalov}. From the picture on
the left, we can find that the subhalos have the power to accelerate
the spaceship to velocity $ 10^{ - 5} c \sim 10^{ - 3} c$ with
reasonable parameters $ S/M = 100m^2 /100ton$. In the first few
hours, the spaceship flies in the saturation region where they will
be accelerated to velocity $ 10^{ - 6} c \sim 10^{ - 4} c$, which
can be understood with the help of Fig.\ref{constdensity} and
Fig.\ref{rsat}. Out of the saturation region, the velocity of the
spaceship can get further boosted by about one order in the $ r^{ -
1}$ density region. Note that the above accelerations take place at
the very center of halo, which is far less than the scale radius $
r_s$. The above results rely on the parameters of spaceship, e.g.
the ratio $ S/M$. If we lower the the weight of spaceship and
increase the area of the engine, the velocity we can achieve will
significantly increase. We specially give the plot on the right for
$ S/M$ which is ten times larger, although the parameters maybe
unreasonable in practice. It shows the the corresponding velocity
increases about ten times. The main reason is the velocity at $ r =
r_{sat}$ increases by ten times, which can be understood with the
Eqn.\ref{conlength1}.

Anyway, the (sub)halos seem difficult to boost the spaceship to
relativistic velocity, because their saturation radius is small
comparing with the required acceleration length $ 10^{ - 4} pc$ in
the Fig.\ref{constdensity}. In the above calculation, we assume
there are no baryonic matter in the halo. The gravity from the DM
halo have negligible effects on the spaceship, even at the
saturation region. Note the saturation density $ \rho _{sat} \sim
10^{19} M_ \odot \cdot kpc^{ - 3}$ is much smaller than the density
of water $ 1g/cm^3 \sim 10^{31} M_ \odot \cdot kpc^{ - 3}$.

However, in case there are baryonic matter in the halo, it may
modify the DM profile. The adiabatic contraction due to dissipating
baryons can steepen the DM profile\cite{Diemand:2009bm}. A more
interesting case is that there is a central black hole in the DM
halo, for example at galactic center. The DM density can become a
dense spike due to accretion by the black hole, assuming adiabatic
growth of the black hole\cite{Gondolo:1999ef}. The annihilations in
the inner regions of the spike set a maximal dark matter density $
\rho _{core}  = \frac{{m_D }}{{\left\langle {\sigma v} \right\rangle
t_{bh} }} \sim 10^{17} M_ \odot   \cdot kpc^{ - 3}$, where $ {m_D }$
is the mass of DM particle, and $ {t_{bh} }$ is the age of black
hole, conservatively $ 10^{10} yr$. And more importantly, the radius
of the core can be as large as $ O(10^{ - 2} pc)$ for inner cusped
model like NFW profile. Recall the Eqn.\ref{conlength}, the required
acceleration length for velocity $0.9c$ is about $ O(10^{ - 2} pc)$
in this case, which means the spaceship can achieve the velocity
close to the speed of light.

\section{Conclusion and discussions}
In this work, we give an example of DM engine using DM annihilation
products as propulsion. The acceleration is proportional to the
velocity, which makes the velocity increase exponentially with time
in the non-relativistic region. The important points for the
acceleration are how dense is the DM density and how large is the
saturation region. The parameters of the spaceship also have great
influence on the results. For example, the velocity will increase if
$ S/M$ increases. We show that the (sub)halos can accelerate the
spaceship to velocity $ 10^{ - 5} c \sim 10^{ - 3} c$ under the
reasonable parameters of spaceship. Moreover, in case there is a
central black hole in the halo, like galactic center, the core
radius of DM can be large enough to accelerate the spaceship close
to the speed of light.

We have used three assumptions in this work. First, we have
assumed static DM for simplicity. But the DM particle may have
velocity as large as $ O(10^{ - 3} c) $. Once we know the velocity
distribution of DM, it can be solved by programming the direction of
the spaceship when speed is low. An analogue in our daily life is airplanes
 work well in both headwind and tailwind. Second, we have assumed the DM particles and the
annihilation products can not pass through the wall of the engine.
For the annihilation products, they may be SM fermions which have
electric charges. Thus we can make them go into certain direction by
the electromagnetic force. The most serious problem comes from DM
which are weakly interacting with matter. Current direct searches of
DM have given stringent bound on cross-section of DM and matter. It
may be difficult using matter to build the containers for the DM,
because the cross-section is very small. However, the dark sector
may be as complex as our baryon world, for example the mirror world.
Thus the material from dark sector may build the container, since
the interactions between particles in dark sector can be large.
Third, the annihilation process is assumed to happen immediately in
the picture C. This is the second serious problem we should pay
attention to. The annihilation speed takes the form, $  A =
\left\langle {\sigma v} \right\rangle \frac{{\rho _{sat}^2
}}{{2m_D^2 }} $. The $ \left\langle {\sigma v} \right\rangle$
 is taken to be the natural scale of the correct thermal relic, which is
 $3 \times 10^{ - 26} cm^3
s^{ - 1}$. One can show that $ A = 2.2 \times 10^{ - 7}
cm^{-3
 } s^{ - 1}$ . However, the number density of the dark matter is $
n_D  = \frac{{\rho _{sat} }}{{m_D }} = 4 \times 10^9 cm^{ - 3} $.
Thus, to make the annihilation process efficient, we have to
compress the volume of the engine to raise the annihilation speed.
Whether it can be achieved in the future is not clear. Nevertheless,
the engine works in the vacuum where the baryonic matter is dilute,
which means we do not need to worry about the pressure from the
baryonic matter.

Sometimes, when looking at the N-body simulation pictures of DM, I
think it may describe the future human transportation in some sense.
In the picture, there are bright big points which stand for large
dense halos, and the dim small points for small sparse halos.
Interestingly, these halos have some common features with the cities
on the Earth. The dense halos can accelerate the spaceship to higher
speed which make it the important nodes for the transportation.
However, the sparse halos can not accelerate the spaceship to very
high speed, so the spaceship there would better go to the nearby
dense halo to get higher speed if its destination is quite far from
the sparse halos. Similarly, if we want to take international
flight, we should go to the nearby big cities. The small cities
usually only have flights to the nearby big cities, but no
international flights. Thus we can understand the dense halos may be
very important nodes in the future transportation, like the big
cities on the Earth.


\appendix*

\section{Dark Matter halo and subhalo profiles}

Based on N-body simulations, the DM distribution can usually be
parameterized as,
\begin{equation}
\rho(r)=\frac{\rho_s}{(r/r_s)^{\gamma}[1+(r/r_s)^{\alpha}]^{(\beta-\gamma)/\alpha}},
\label{nfwmoore}
\end{equation}
where $\rho_s$ and $r_s$ are the scale density and the scale radius
parameters respectively. The parameters $(\alpha,\beta,\gamma)$ are
$(1,3,1)$ for NFW profile. In this appendix, we briefly introduce
how $r_s$ and $\rho_s$ are calculated. The two parameters can be
determined once we know the (sub)halo mass $M_v$ and the
concentration parameter $c_v$ which depends on the specific
concentration model. The calculations are following the method in
Ref.\cite{Lavalle:1900wn}. In the appendix of Ref.\cite{Liu:2008ci},
we have shown how to determine the $r_s$ in detail.

For the NFW profile, the scale radius $r_s$ is

\begin{equation} r_s^{nfw}=\frac{r_v(M_v)}{c_v(M_v)}. \label{rsnfweqn}\end{equation}

where $r_v$ is the virial radius of the subhalo which is often
approximated as the radius within which the average density is
greater, by a specific factor $\Delta=200$, than the critical
density of the Universe $\rho_c=139 M_{\odot} kpc^{-3}$ ($M_{\odot}$
is mass of the Sun). Thus $r_v$ can be expressed as $
 r_v = \left( {\frac{{M_v }}{{(4\pi
/3)\Delta \rho _c }}} \right)^{1/3} $. The $c_v$ is the
concentration parameter of the subhalo which is determined by the
subhalo mass $M_v$ and concentration model. We use the same method
as Ref.\cite{Lavalle:1900wn} which adopts two concentration models,
which are ENS01\cite{eke01}  and B01\cite{bullock01}. In the
Ref.\cite{Lavalle:1900wn}, the $c_v$ is fitted in a polynomial form
as
\begin{equation}
\ln(c_v)=\sum_{i=0}^4C_i\times\left[\ln\frac{M_v}{M_{\odot}}\right]^i,
\label{lncv}
\end{equation}
where $C_i=\{3.14,-0.018,-4.06\times10^{-4},0,0\}$ and
$\{4.34,-0.0384,-3.91\times10^{-4},-2.2\times10^{-6},
-5.5\times10^{-7}\}$ for ENS01 and B01 model respectively.

\begin{figure}[!htb]
\begin{center}
\includegraphics[width=3.4in,angle=0]{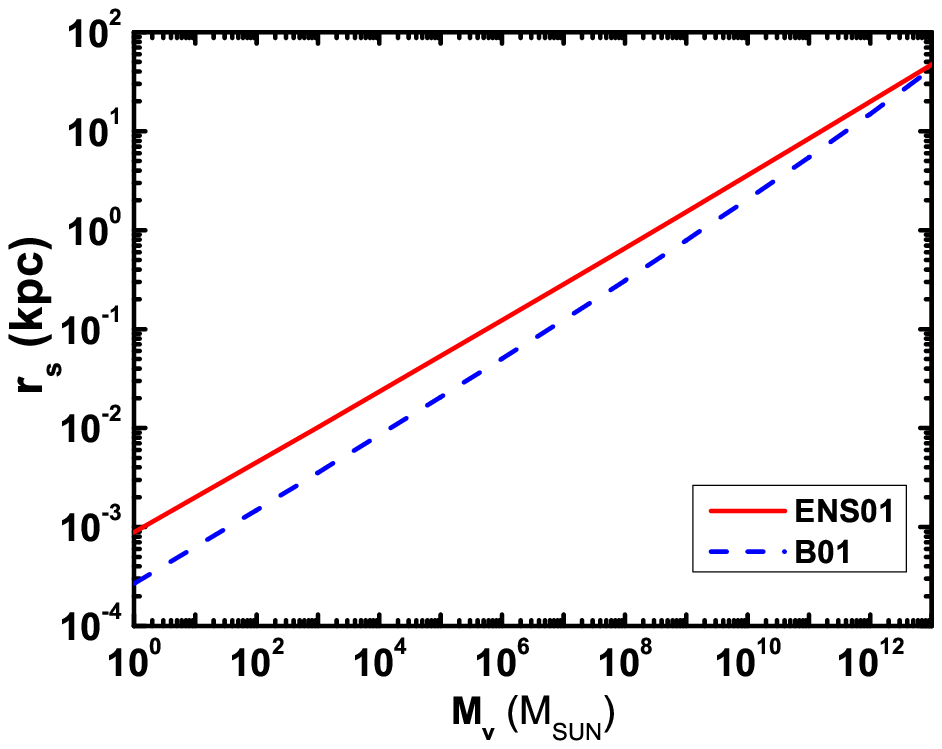}%
\includegraphics[width=3.4in,angle=0]{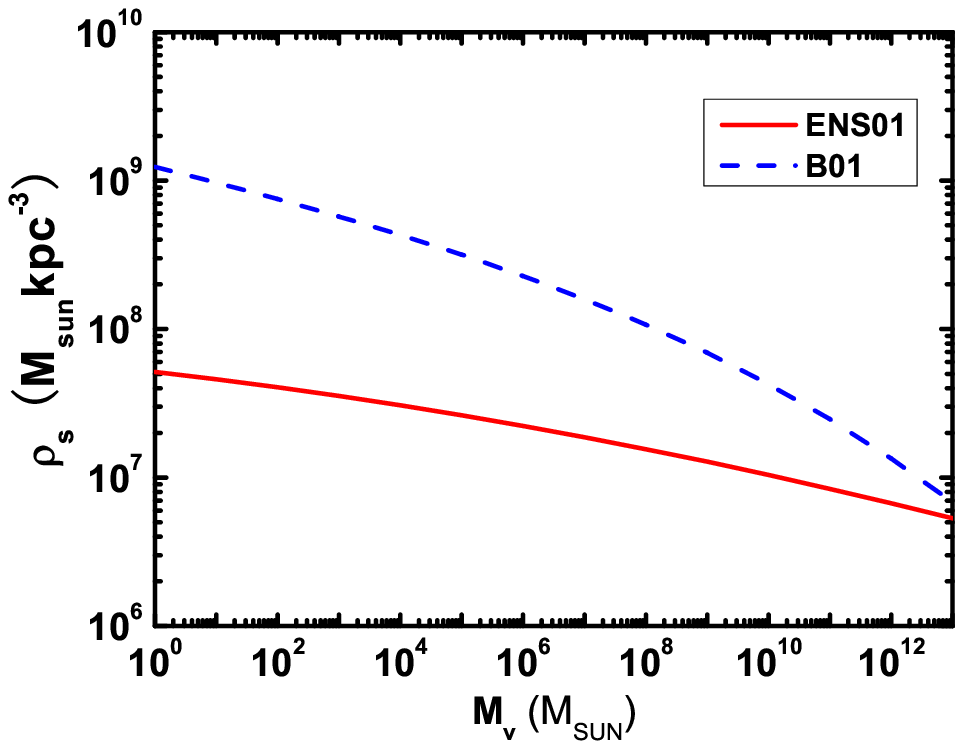}
\caption{The scale radius $r_s$ and the scale density $\rho_s$ as a
function of (sub)halo mass $M_{v}$. This plot assumes the (sub)halos
have NFW profile. }\label{rsrhos}
\end{center}
\end{figure}

The density scale $\rho_s$ can be determined by the mass relation
$\int\rho_s(r) {\rm d}V=M_v$. One can get the scale density
$\rho_s$,
\begin{equation}
\rho _s^{nfw}  = M_v /[4\pi r_s^3 A(c_v )],
\end{equation}

where $ A(c_v ) \equiv In(1 + c_v ) - c_v /(1 + c_v )$. In
Fig.\ref{rsrhos}, the scale radius $r_s$ and the scale density
$\rho_s$ are plotted as a function of subhalo mass $M_{v}$ to show
how large and how dense the subhalos are. We can see the scale
radius $r_s$ is quite large which shows the acceleration is mostly
done in the $ r \ll r_s$ region of the (sub)halo.

\end{document}